\documentclass{article}

\usepackage{arxiv}

\usepackage[utf8]{inputenc} 
\usepackage[T1]{fontenc}    
\usepackage{hyperref}       
\usepackage{url}            
\usepackage{booktabs}       
\usepackage{amsfonts}       
\usepackage{nicefrac}       
\usepackage{microtype}      
\usepackage{lipsum}
\usepackage{bm}
\usepackage{amsmath, amssymb}
\usepackage{mathptmx}
\usepackage{mathrsfs}
\usepackage{graphicx}
\usepackage{multirow}
\usepackage{comment}

\title{paradigm shift through the integration of \\ physical methodology and data science}

\author{
 Takashi Miyamoto\thanks{Corresponding Author} \thanks{Also affiliated with German Research Center for Artificial Intelligence, Kaiserslautern, Germany} \\
  Department of Civil and Environmental Engineering\\
  University of Yamanashi\\
  Yamanashi, Japan\\
  \texttt{tmiyamoto@yamanashi.ac.jp}
  
}

\begin{document}
\maketitle
\begin{abstract}
Data science methodologies, which have undergone significant developments recently, provide flexible representational performance and fast computational means to address the challenges faced by traditional scientific methodologies while revealing unprecedented challenges such as the interpretability of computations and the demand for extrapolative predictions on the amount of data. Methods that integrate traditional physical and data science methodologies are new methods of mathematical analysis that complement both methodologies and are being studied in various scientific fields. This paper highlights the significance and importance of such integrated methods from the viewpoint of scientific theory. Additionally, a comprehensive survey of specific methods and applications are conducted, and the current state of the art in relevant research fields are summarized.
\end{abstract}

\section{Introduction}
Data science technology, which extracts and generates useful knowledge from data obtained in various forms, has made significant progress recently as a methodology leading to a new paradigm in science\cite{Shapere1964-zp} and has been applied in various scientific fields such as informatics\cite{Deng2014-ru,LeCun2015-ki}, biology\cite{Baxevanis2020-rd}, and materials science\cite{Agrawal2016-tb}. Under the digital twin\cite{Tao2019-nn}, which is proposed as a core system for the next generation of society, data science is expected to play an increasingly important role in basic research and in the real world.

However, the trend toward the application of data science methods has also brought to light issues common to many application fields. The lack of data because of the lack of opportunities to observe rare events\cite{Qi2020-lk} and the black-box nature of the computational process highlighted using mainstream machine learning methods\cite{Castelvecchi2016-rv} are typical examples of the major challenges in data science. 
Additionally, many problems in science and engineering require the discovery of laws that describe various phenomena in a unified and abstract way, and the prediction and control of phenomena based on these laws. However, realizing the elucidation of such universal laws and extrapolative prediction using data science methods, which are mainly based on techniques of interpolation of training data, is a significant challenge, and several studies are being conducted\cite{Sahoo2018-wo,Raissi2018-ha}.

In contrast to these issues with the application of data science methods, scientific methodologies developed prior to the rise of data science have sought to discover and apply universal laws based on deductive reasoning. Additionally, with computer advancements, it is now possible to predict even complex phenomena with high accuracy through numerical analysis of the laws described in the form of governing equations. However, even under such scientific methodologies, there are cases where the computational cost of performing calculations based on principles is prohibitively high, and the accuracy of a physical model always depends on the validity of the assumptions and approximations used in the formulation. In such cases, the main research topics remain speeding up the computation\cite{Kindratenko2011-hf} and improving the modeling accuracy\cite{Khain2015-ca}.

Data science methodology, which builds computational models using data, and traditional scientific methodology, which builds computational models using deductive reasoning, are contradictory. Recently, methods have been proposed that are intended to solve the problems of both methodologies by combining the speed and flexibility of data science methodologies with transparency of computational processes and affinity with natural principles of conventional scientific methodologies. Such integrated methodologies extend the framework of scientific analysis of phenomena to include various scientific and engineering applications, and are being studied in various interdisciplinary areas.

Therefore, this study discusses the significance and importance of methods for analyzing phenomena that integrate conventional scientific and data scientific methodologies from the perspective of the scientific paradigm shift, and examines the advantages of such a method compared to the use of each methodology alone. Additionally, specific integrated methods and applications are surveyed and categorized, and current state of the art in the research field is summarized.

\section{Integrated methods as a scientific paradigm}
\subsection{Paradigm in science}
In science, a paradigm is defined as a widely accepted scientific achievement that provides the scientific community with a normative problem formulation and solution for a given period\cite{Shapere1964-zp}.
According to Kuhn's definition, the current scientific paradigm is "a widely accepted normative methodology for a period that defines how scientific problems are set and solved." Gray\cite{Hey2009-vr} summarized recent scientific paradigms and describes the evolution of methodologies in the order of empirical, theoretical, and computational sciences with the development of scientific knowledge and methods, as shown in the Table \ref{tab:3paradigm}.

\begin{table*}[!ht]
    \centering
    \caption{Recent three science paradigms\cite{Hey2009-vr}}
    \label{tab:3paradigm}
    \begin{tabular}{lll}
    \hline \hline
         Paradigm & Period & Explanation  \\
         \hline
         Empirical science & From a thousand years ago & The data acquisition and empirical description of phenomena \\
         Theoretical science & Last few hundred years & Modeling and generalization through deductive theory \\
         Computational science & Last few decades & Numerical simulation of complex phenomena \\
         \hline \hline
    \end{tabular}
\end{table*}

A paradigm shift does not imply that a new methodology replaces an old one, but rather that the framework of analysis has been extended through the coordination and integration of different methodologies. As an example of an analytical method in modern natural science, the process is as follows:

\begin{enumerate}
    \item Collect data on a target phenomenon through observation or experimentation (empirical science)
    \item Represent the laws behind the data through governing equations (theoretical science)
    \item Solve the governing equations through discretization and numerical computation (computational science)
\end{enumerate}

\begin{figure}[h]
    \centering
    \includegraphics[width=130mm]{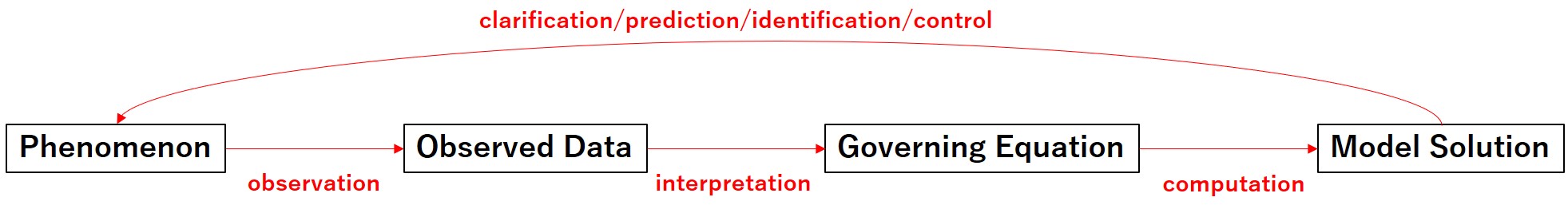}
    \caption{The process of knowledge discovery and problem solving in the modern scientific paradigm}
    \label{fig:sciencemethod}
\end{figure}

\begin{figure}[h]
    \centering
    \includegraphics[width=130mm]{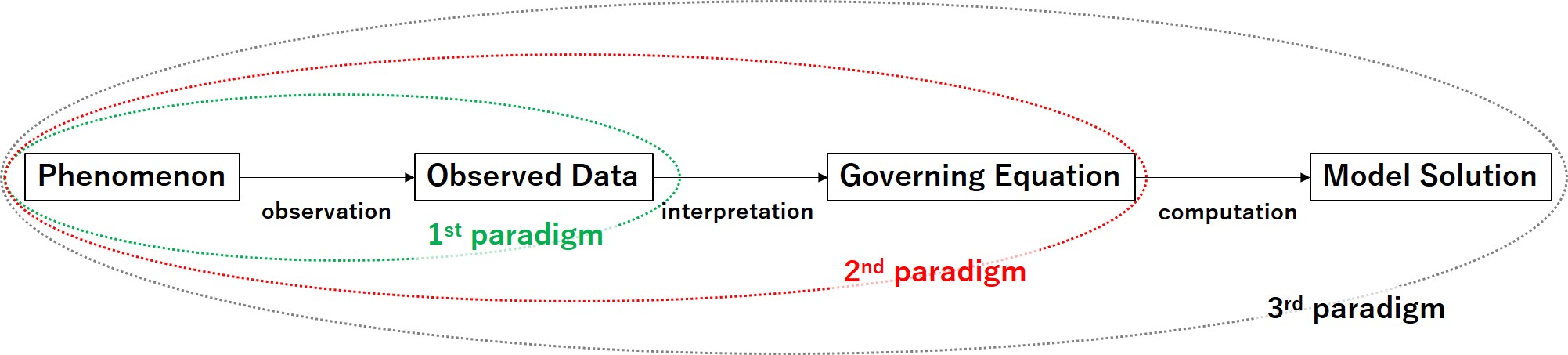}
    \caption{The paradigm shift as a expansion of analysis framework}
    \label{fig:3paradigms}
\end{figure}

We can clarify, predict, identify, and control the original phenomenon, leading to the discovery of scientific knowledge and the resolution of problems (Figure \ref{fig:sciencemethod}). This process is composed of methods belonging to different paradigms, such as the collection of observational data, modeling with governing equations, discretization and computation, which complement each other's domain. In other words, a paradigm shift represents an update of the normative methodology of science in a way that encompasses the previous methods, while expanding the framework for knowledge discovery and problem solving (Figure \ref{fig:3paradigms}).

\subsection{Paradigm shift led by data science}
Gray\cite{Hey2009-vr} further points out that science is transitioning to the fourth paradigm with new technologies to process the data generated from observations and numerical simulations, as the volume and complexity of data becomes too large for humans to process. The domain of these technologies is called data-intensive science, and the following two technological domains are indicated as directions:
\begin{itemize}
    \item Informatics: technologies for observing and collecting data and analyzing it for meaningful information
    \item Computation: technology for simulating the properties and behavior of an object 
\end{itemize}

\begin{figure}[h]
    \centering
    \includegraphics[width=130mm]{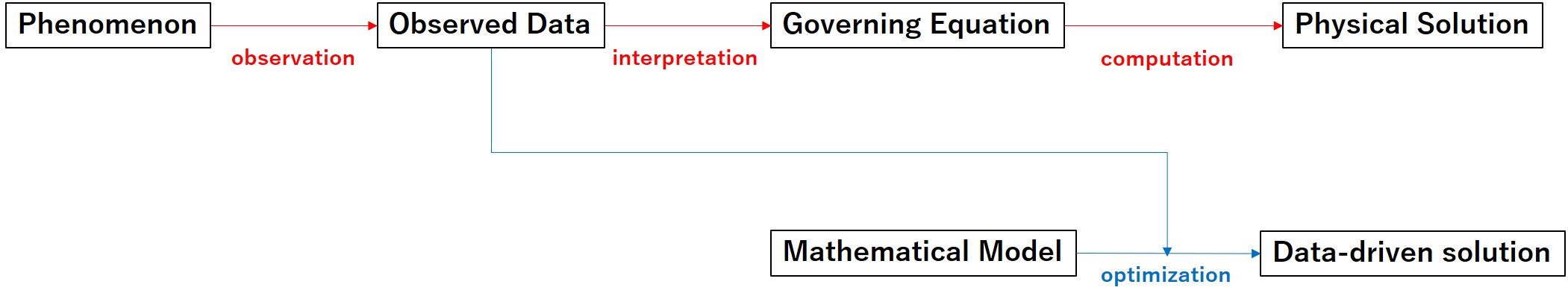}
    \caption{Comparison of physical and data science methodologies for modeling}
    \label{fig:ml}
\end{figure}

Machine learning techniques, for example, that fall under the domain of computation, can be explained in contrast to traditional methodologies. The traditional paradigm (physical methodology) model phenomena using governing equations derived from deductive reasoning of observed data, and their solutions are obtained using computational science methods. However, machine learning techniques use mathematical models that are optimized in a data-driven manner, rather than being modeled by deductively derived equations (Figure \ref{fig:ml}). Such a data science methodology for modeling such phenomena provides a flexible means of modeling and solving phenomena that are difficult to describe explicitly using the governing equations. The great success of recent data-driven models, such as deep learning, demonstrates the effectiveness of data science methodologies in various fields.

\subsection{Integrating traditional scientific paradigms with data science}
\begin{figure}[h]
    \centering
    \includegraphics[width=130mm]{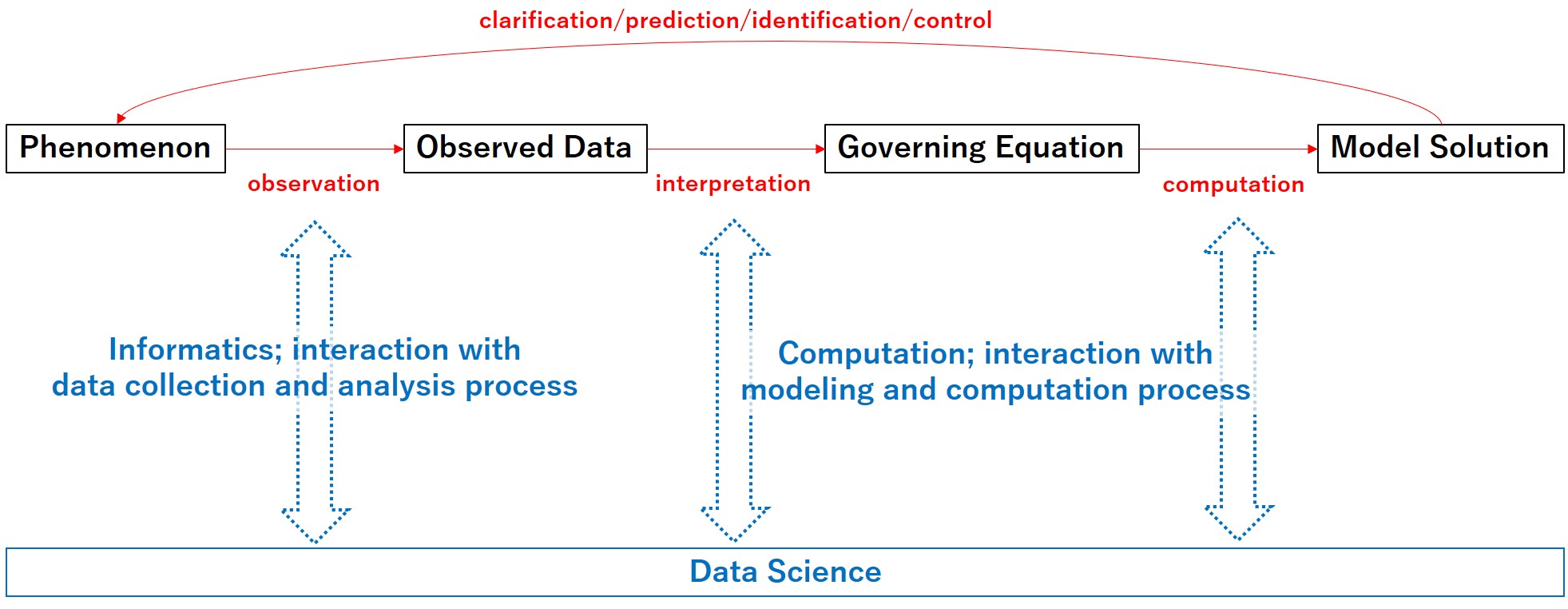}
    \caption{Relationship between traditional methodologies and data science}
    \label{fig:dscience}
\end{figure}

The methodologies offered by data science go beyond the automatic modeling of phenomena as described above.
Brunton et al.\cite{Brunton2019-ke} state that data science "continues to revolutionize the way we model, predict, and control phenomena," and collectively refer to the workings of data science as "data-driven knowledge discovery.” The essence of data science can be understood as the interaction between knowledge discovery and problem-solving processes in the traditional scientific paradigm, as well as the two technical domain categories in data science presented in \cite{Hey2009-vr} (Figure \ref{fig:dscience}).

\begin{figure}[h]
    \centering
    \includegraphics[width=130mm]{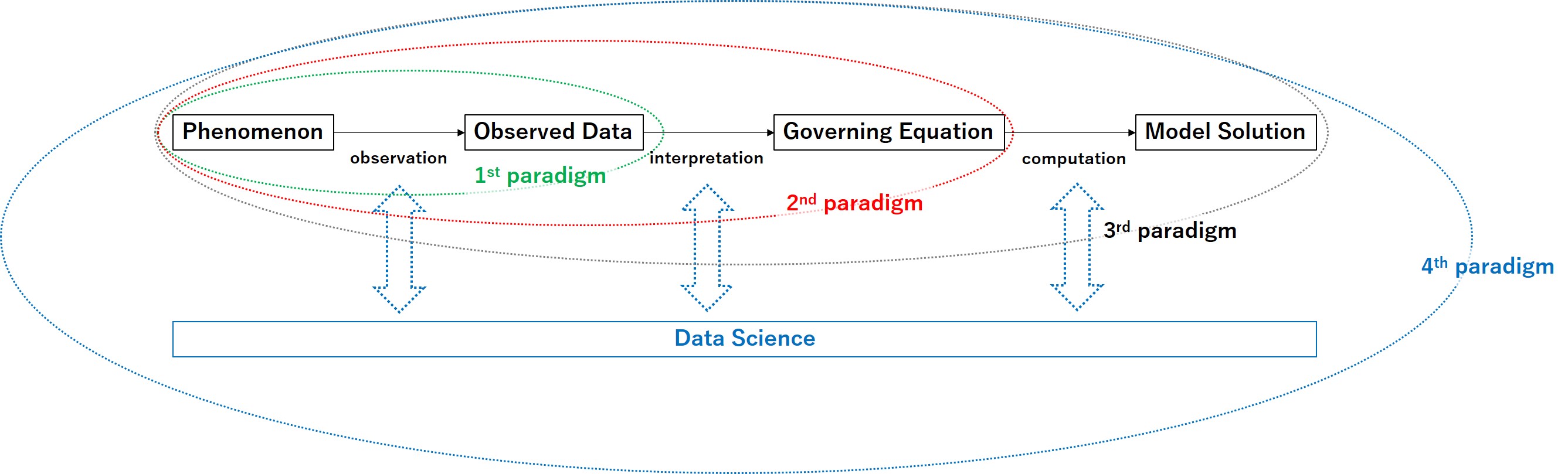}
    \caption{Paradigm shift led by collaboration between traditional methodologies and data science}
    \label{fig:4thparadigm}
\end{figure}

Under the light of the previous review, we can argue that data science is more than just a replacement for the existing analytical process shown in Figure \ref{fig:ml}, but also a technology that leads to a paradigm shift in the cooperative relationship with the conventional methodology shown in Figure \ref{fig:dscience}. Based on the history of scientific paradigm shifts, it can be expected that the technology that connects traditional paradigmatic methods with data science methods will function effectively as a methodology that expands the means of knowledge discovery and problem solving (Figure \ref{fig:4thparadigm}).

\section{Purpose of integrating physics and data science}
So far, we have discussed the significance and importance of integrating traditional scientific paradigms and data science from a science theory perspective. In fact, recently, several approaches that integrate the two methodologies have been proposed and shown to be effective in solving problems that may arise when conventional physical and data science methodologies are applied independently, particularly in the modeling process of phenomena.

Therefore, in this and the following sections, we review integrated methods of physical and data science methodologies for modeling phenomena, and summarize the purposes and types of methods used. There are already excellent compilations of information on such integrated methods in related works\cite{Willard2020-ag,Rai2020-mo}. However, these studies have been reviewed mainly from the viewpoint of extending data science methodologies; in this study, we will reorganize and discuss the information by referring to these results and adding the viewpoint of extending physical methodologies.

Physical and data science modeling methodologies have conflicting advantages and challenges. The basic idea of integrated physics and data-driven methodologies is to overcome these challenges by using each methodology in a complementary manner. The advantages of the integrated method over physical and data science methods are discussed in detail below.

\subsection{Advantages over physical methods}
\subsubsection{Increasing the accuracy of the model}
Physical models are widely applied in science and engineering, and have been used with great success to date. however, the governing equations used to describe phenomena are generally developed through assumptions, simplifications, and approximations, whereas parameterization schemes\cite{Stephens1984-qs} are used to describe phenomena that are difficult to develop directly in complex physical processes or on a sub-grid scale. The limitations in the representation capability and predictive accuracy of physical models owing to these simplifications and approximations continue to be an issue in several scientific fields\cite{Gupta2014-bb,Ghil2020-sh}.

However, data science models, such as machine learning, can accurately reproduce the input-output relationships of highly nonlinear data under the assumption that there are sufficient training data. Therefore, research is being conducted to improve the accuracy of the overall model by supplementing the differences between physical models and real phenomena caused by simplification and approximation with data science models with flexible representation capabilities.
In such efforts, bias terms\cite{Xu2015-ej} and parameterization terms\cite{Brenowitz2018-wj} representing errors between physical model predictions and actual observed data, as well as some physical quantities in coupled physical processes\cite{Hamilton2017-tt}, are output to the data-driven models to improve the final prediction accuracy.

\subsubsection{Faster computation}
With the progress of computer science\cite{Shimizu2020-ve} and high-performance computing methods\cite{Kindratenko2011-hf}, the spatiotemporal scale and resolution of computations using physical models have continued to improve. However, computational cost is a bottleneck in applications such as the reproduction of physical phenomena that require global-scale computations\cite{Massoud2019-rc} and engineering applications that require real-time simulations\cite{Zanchetta2020-hr}. Furthermore, recently, there has been a growing demand for performing these computations several times to quantify the uncertainty\cite{Wu2020-sp}.

Various data science approaches have been used to solve the computational cost problem in physical models. The replacement of a physical model by a faster data science model trained on physical simulation results is called a surrogate model. Surrogate models have been used for large model ensemble computations\cite{Tran2020-wq}, Monte Carlo simulations\cite{Su2017-yk} and inverse problems\cite{Bilicz2012-xf}. In some cases, the convergence of iterative algorithms has been improved using the output of surrogate models as initial values for iterative algorithms in physical models\cite{Gafeira2021-ad}.

Super-resolution technology, which improves the spatiotemporal resolution of data through nonlinear interpolation, is used to improve the resolution of observed data and the output values of physical models\cite{Vandal2017-vh} as an approach to speed up the computation different from surrogate models. Additionally, reduced-order models, which reduce the computational dimensionality by extracting the main modes inherent in complex physical phenomena for prediction, have been applied in fields such as fluid engineering\cite{Lassila2014-ow}. In addition to the use of modes obtained through mathematical procedures\cite{Schmid2011-rw,Ravindran2000-op}, data-driven mode acquisition has recently been studied\cite{Wang2018-hn,Kaiser2017-yo}.

\subsection{Advantages over data science methods}
\subsubsection{physical consistency}
Data science models have a high representation capability to reproduce the nonlinear input-output relationships of data on various phenomena\cite{LeCun2015-ki}. However, most mainstream machine learning models have a black box nature in their computational process, and the computational results generally do not satisfy physical laws such as conservation laws. This may lead to the problem of interpretability of the computation process and results as well as the consistency of data science models with physical knowledge.

To solve this problem, methods have been developed to ensure the physical consistency of machine learning models by imposing constraints based on physical knowledge on the output results and computational processes of machine learning models. By adding a physical condition to the loss function as a regularization term, the output of the machine learning model satisfies the physical condition and restricts the solution space, improving the learning efficiency from small training data\cite{Karpatne2017-uo}. Similarly, by incorporating physical knowledge into the architecture of machine learning models, the interpretability and physical consistency of the computational process, learning efficiency, and prediction performance can be improved\cite{Ling2016-xo}.

\subsubsection{Improving extrapolation capability}
A data science model trained on enough data shows generalization performance and high prediction performance, even for untrained data. However, the generalization performance of data science models is realized as interpolation in the feature space\cite{Rifai2011-kw}, and extrapolative prediction using input data with completely different characteristics than the training data is considered difficult. Therefore, to build a model with high prediction performance for various input data, it is necessary to prepare training data that cover all possible events; however, in reality, it is difficult to collect such comprehensive data when targeting rare or exceptional phenomena.

In contrast, physical models do not require training data and can be universally applied during predictions based on the laws expressed by the governing equations. Therefore, some methods have been proposed to provide extrapolation capability to data science models by incorporating physical knowledge in the model design and output\cite{Singh2019-in}. Extrapolation performance in data science models means that the model might show high prediction performance even with a small amount of training data or for phenomena not present in the training data, which are expected to expand the applicability of the model.

\section{Integrated methods of physics and data science}
\begin{figure}
    \centering
    \includegraphics[width=130mm]{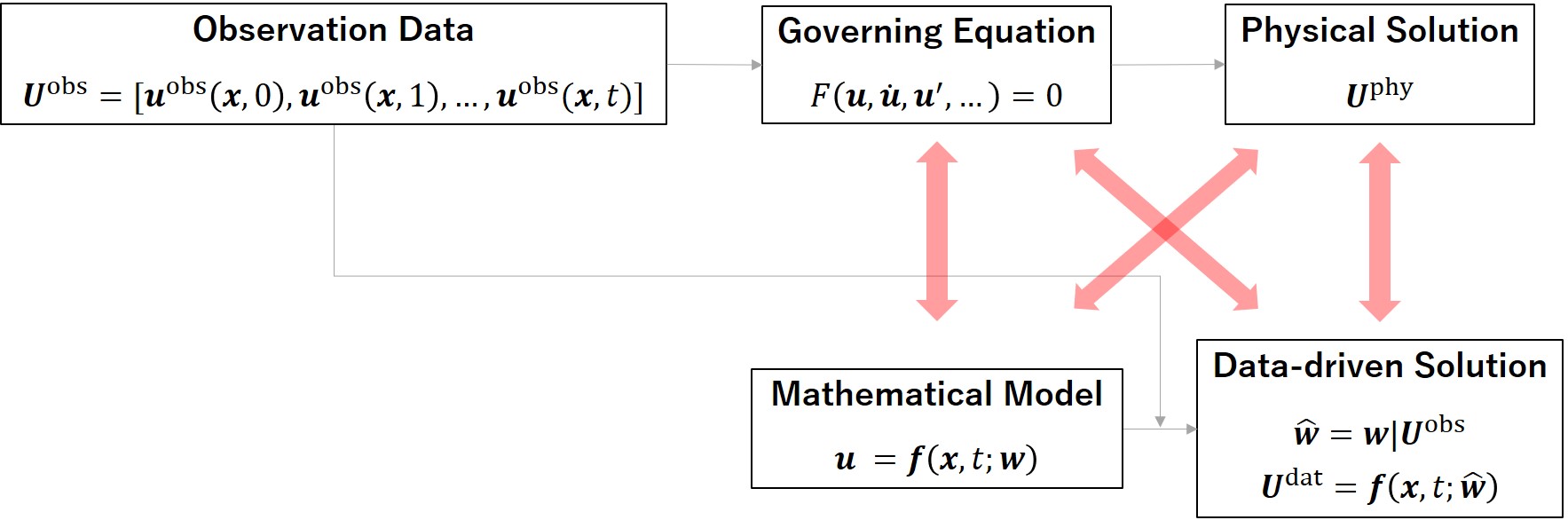}
    \caption{Integrating the Modeling Process}
    \label{fig:modeling}
\end{figure}

To obtain the advantages discussed in the previous section, various methods that integrate physical knowledge and models with data science models in various ways have been proposed. In this study, we describe these methodologies by making a taxonomy in terms of where the integration occurs in the modeling process based on Figure \ref{fig:modeling}.
This section assumes the problem of modeling a physical quantity $\bm{u}$ distributed across space-time, where $\bm{x}$ and $\bm{t}$ are its spatial coordinates and time. $\bm{U}^{\text{obs}}=[\bm{u}^{\text{obs}}(\bm{x},t)]$, $\bm{U}^{\text{phy}}$ and $\bm{U}^{\text{dat}}$ are the observed data, physical model predictions, and data-driven model predictions for $\bm{u}$, respectively.
$F(\bm{u}, \dot{\bm{u}}, \bm{u}',\cdots)$ represents an arbitrary operator, and thus $F(\bm{u}, \dot{\bm{u}}, \bm{u}',\cdots)=0$ represents a governing equation such as a partial differential equation.
$\bm{f}(\bm{x},t;\bm{w})$ represents the mathematical model that predicts $\bm{u}$ from $\bm{x},t$, with parameter $\bm{w}$ optimized by learning the observed data $\bm{U}^{\text{obs}}$. Note that The notation $x|y$ implies that the value of $x$ is determined by a given $y$.

\begin{figure}
\centering
\begin{tabular}{ll}
     \includegraphics[height=22mm]{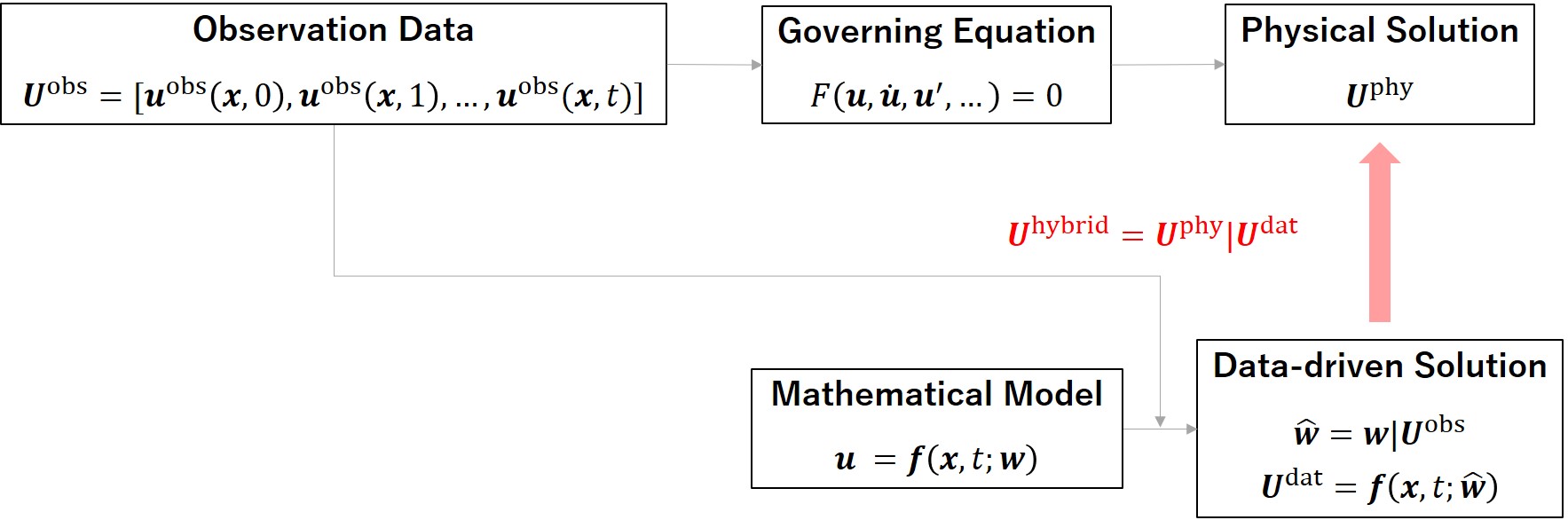} & 
     \includegraphics[height=22mm]{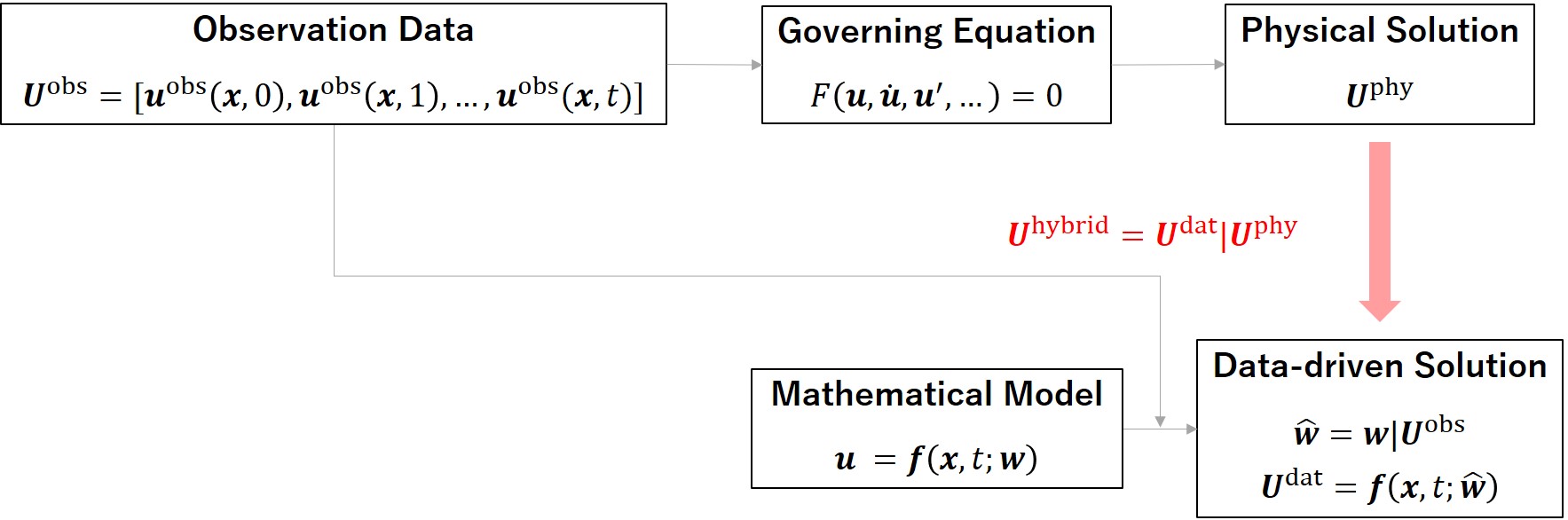} \\
     (a)Solving physical models using data-driven solutions &
     (b)Optimizing data-driven Models using physical solutions \\ \\ \\ \\
     \includegraphics[height=22mm]{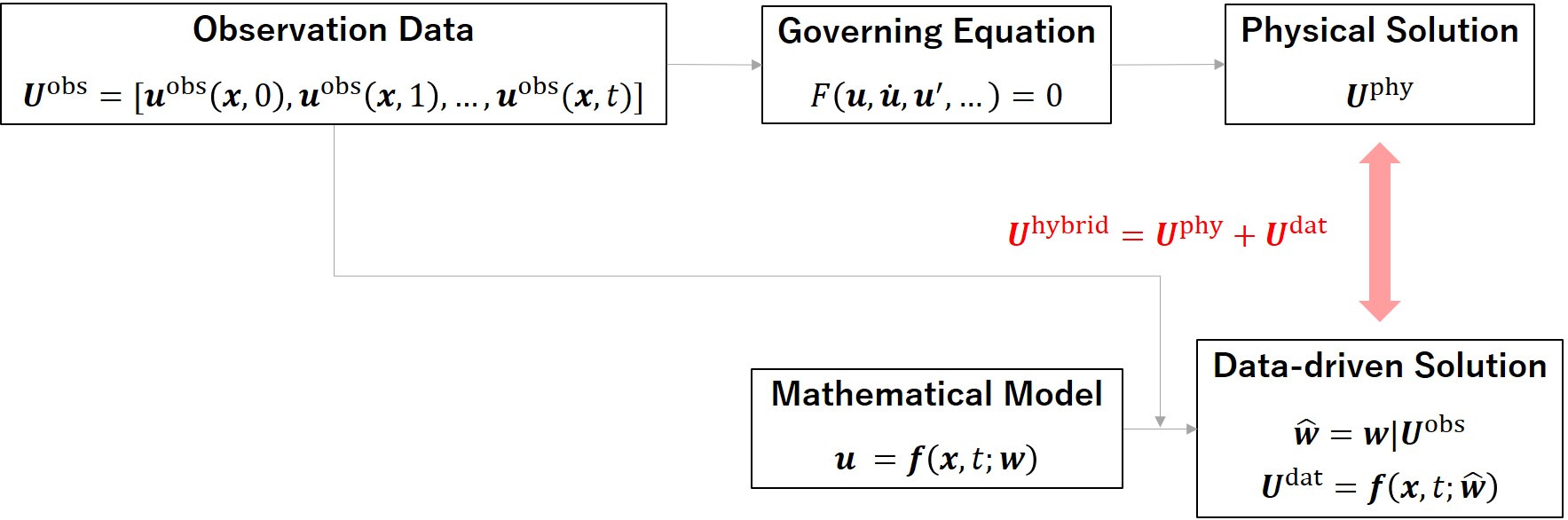} & 
     \includegraphics[height=22mm]{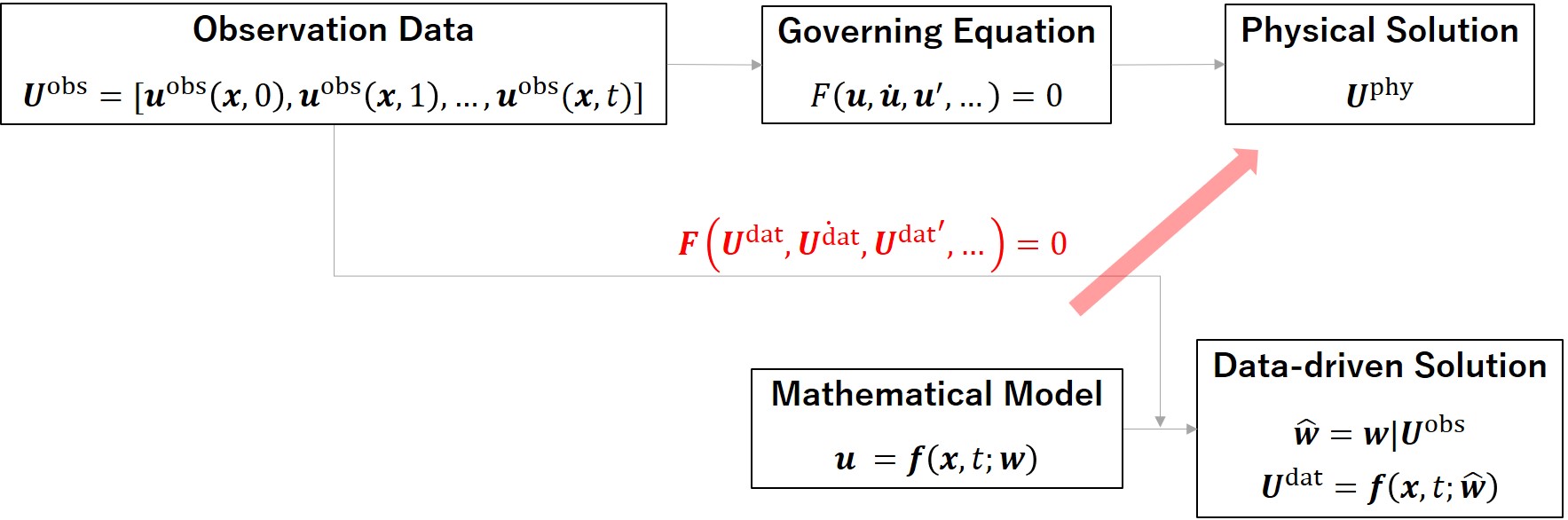} \\
     (c)Hybrid of physical and data-driven solutions &
     (d)Solving the governing equations using data-driven models \\ \\ \\ \\
     \includegraphics[height=22mm]{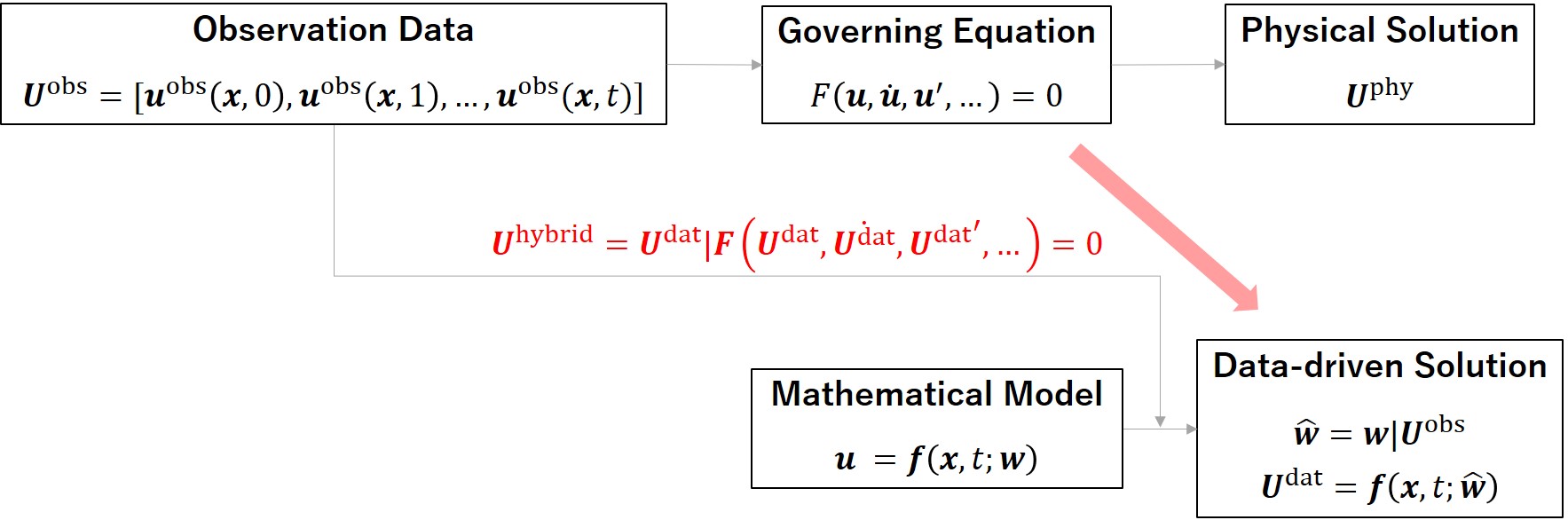} & 
     \includegraphics[height=22mm]{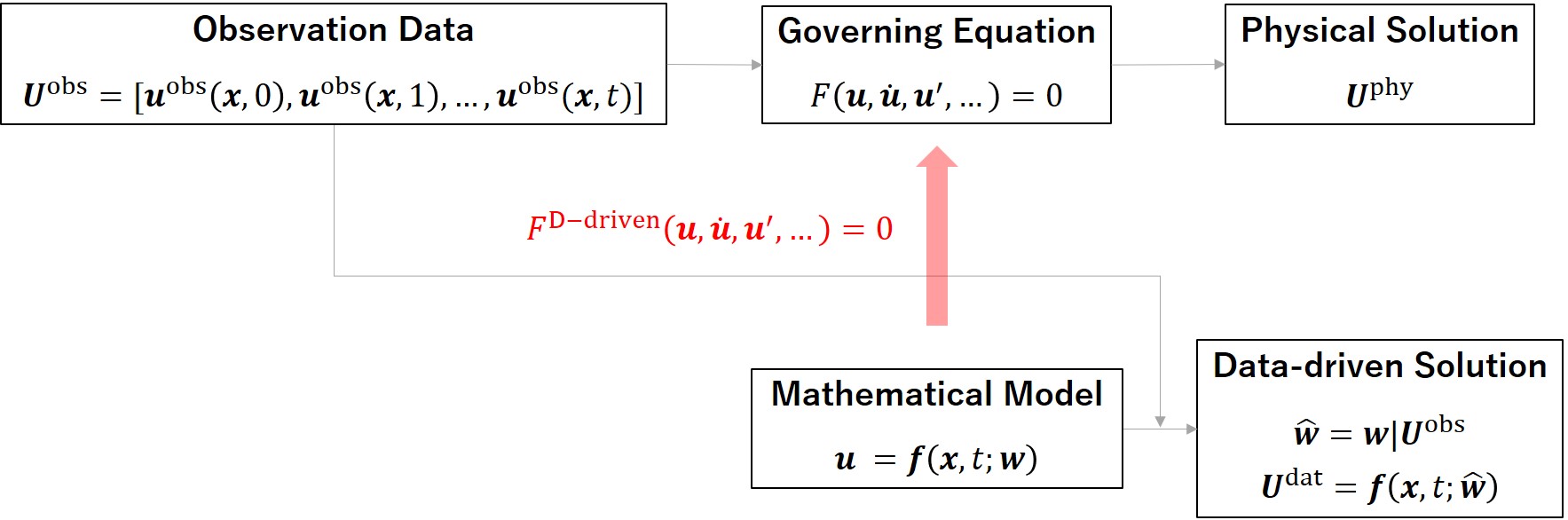} \\
     (e)Physics-based loss functions &
     (f)Data-driven construction of governing equations \\ \\ \\ \\
     \includegraphics[height=22mm]{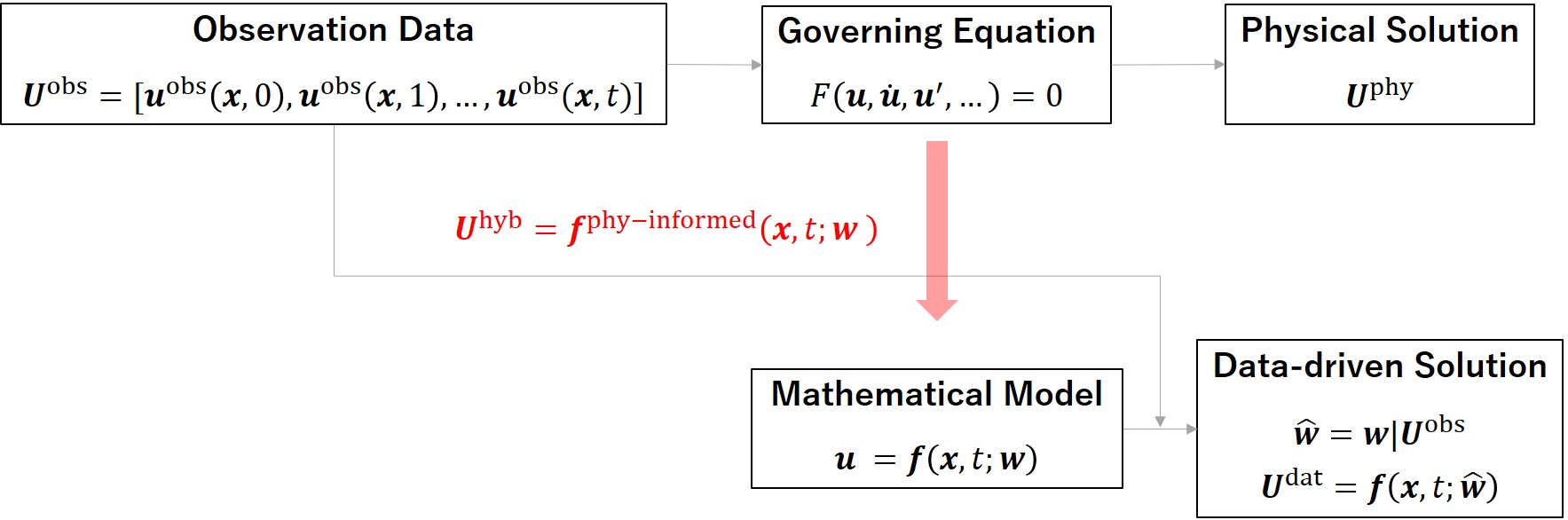} & 
     \includegraphics[height=22mm]{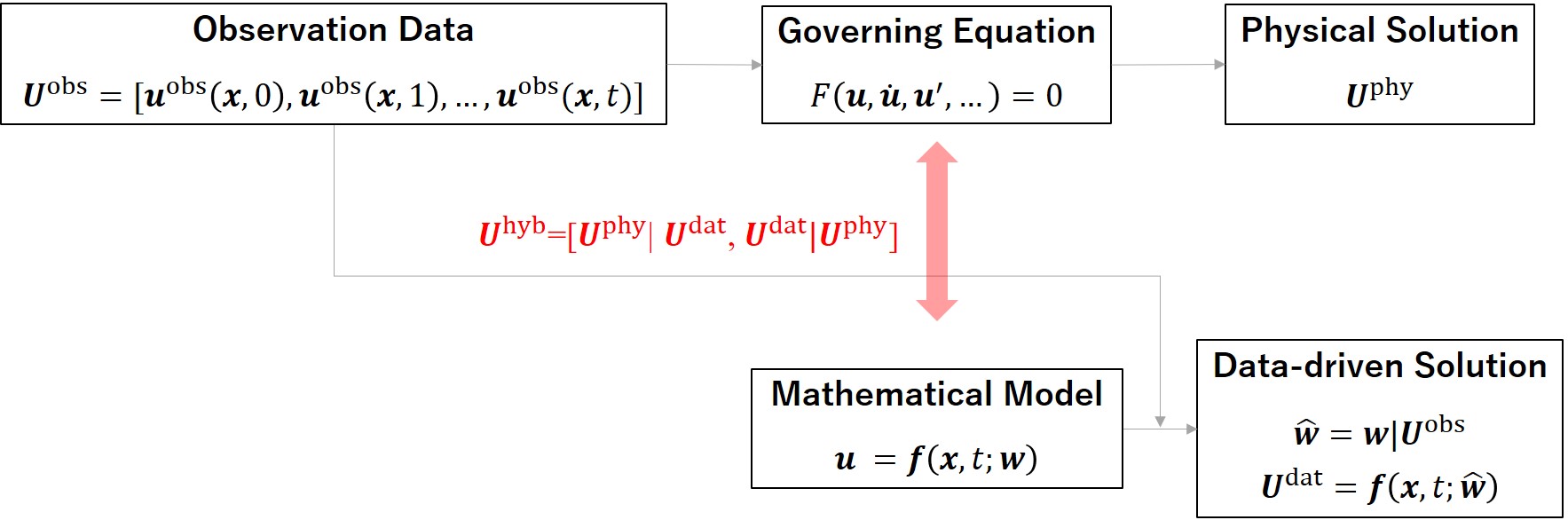} \\
     (g)Reflecting physical knowledge in model architectures &
     (h)Coupled analysis of physical and data-driven models
\end{tabular}
\caption{A taxonomy of methods for integrating physics and data science}
\label{fig:type}
\end{figure}

Each subsection 4.1-4.8 will be explained in relation to with (a)-(h) in Figure \ref{fig:type}.

\subsection{Solving physical models using data-driven solutions}
The process of solving a physical model consists of pre-processing, such as element decomposition, the main computation to obtain the solution, and post-processing for visualization. The recent demand for large-scale scientific computations and the increase in the computational cost of these processes has become a major bottleneck. Therefore, methods for obtaining physical model solutions by speeding up the computational process using data-driven solutions have also been proposed. The solution $\bm{U}^{\text{hybrid}}$ of the integrated method obtained here can be formally expressed as follows:

\begin{equation}
U^{\text{hybrid}}=U^{\text{phy}}|U^{\text{dat}}    
\end{equation}

The quality of the element decomposition is important in discretization methods such as the finite element method for analyzing arbitrary shapes with high accuracy, but the generation of high-quality elements requires a large amount of human labor and computational cost proportional to the size of the target analysis and the required resolution. Zhang et al.\cite{Zhang2020-uk} proposed a method to automate such element generation using a machine learning model and a conventional solver for the solution process after element decomposition. Because element generation is a problem where both the quality of the generated elements and the computational speed are important, it is an area where data-driven methods with flexible modeling performance and high speed can be expected to work effectively.

Additionally, Gafeira et al.\cite{Gafeira2021-ad} demonstrated that in the inverse problem of estimating the state of the solar atmosphere from observed data of the magnetic field produced by the sun, the convergence of the algorithm is improved using the output solution of the machine learning model as the initial value of the iterative algorithm. A similar approach using partially data-driven solutions in a physical solution algorithm was also adopted by Ichimura et al.\cite{Ichimura2018-xp} in a forward problem in solid mechanics. The data-driven method is partially used to improve the convergence speed of the algorithm in these studies, and the entire computational process follows a physical methodology, which can guarantee the physical consistency of the analysis results while taking advantage of the speed of the data-driven model.

\subsection{Optimizing data-driven Models using physical solutions}
Deep learning, which has recently become a mainstream machine learning method, is an approach that uses neural networks with a complex layered structure to achieve flexible expressive capabilities. The optimization of such a neural network requires a large amount of training data corresponding to the number of parameters in the model. However, because of the limitations of data observation and the costs associated with labeling, it is often difficult to prepare such data in practice. Therefore, recently, many approaches have been adopted to use virtual solutions obtained from physical simulations as training data for data-driven models. The solution $\bm{U}^{\text{hybrid}}$ of the integration method obtained here is expressed as follows.

\begin{equation}
\bm{U}^{\text{hybrid}}=\bm{U}^{\text{dat}}|\bm{U}^{\text{phy}}    
\end{equation}

In the research field of civil engineering, a benchmark dataset for image recognition of structures was published\cite{cg}, which uses computer graphics to create images of civil engineering structures damaged by disasters or aging. In the field of hydrology, Jia et al.\cite{Jia2021-os} demonstrated that a neural network can be pre-trained with physical simulation results and then fine-tuned with a small amount of real data to develop a water temperature prediction model with high performance. Furthermore, Read et al.\cite{Read2019-mp} demonstrated that the same approach can also be used to obtain extrapolative prediction performance under different conditions from the training data. Such an approach of adding physical virtual data to the training data has been studied in various fields\cite{Hurtado2018-dw,Shah2018-gm} and is considered an effective method for extrapolating and supplementing the training data with a physical model when the number of real data is limited.

\subsection{Hybrid of physical and data-driven solutions}
Owing to the effects of approximations, assumptions, and simplifications in modeling using governing equations, errors (biases) can occur between the predictions of physical models and the values of real phenomena. The hybridization of physical and data-driven solutions is a method for obtaining more accurate predictions than the physical model alone because the data-driven model learns the bias between physical solutions and real phenomena. The solution using such a hybridization method can be expressed as follows.

\begin{equation}
    \bm{U}^{\text{hybrid}}=\bm{U}^{\text{phy}}+\bm{U}^{\text{dat}}
\end{equation}

Hybrid methods such as bias correction have been widely used in meteorology\cite{Cho2020-co}, and Kubo et al. \cite{Kubo2020-fg} used a similar approach to predict the intensity of ground motion during earthquakes. The use of data-driven models to correct the physical solutions was previously more widely used\cite{Thompson1994-bh}; however, as the representational capabilities of data-driven models have improved, the effectiveness of this approach has increased.

\subsection{Solving the governing equations using data-driven models}
Algorithms for solving the forward and inverse analyses described by the governing equations require computational costs proportional to the scale of the computation. However, as mentioned in the previous section, it is impractical to accept such computational costs in real problems. Therefore, research has been conducted to obtain solutions to the governing equations quickly using models such as neural networks as solvers. The solution obtained from this approach can be expressed as a data-driven solution that satisfies the governing equations as follows:

\begin{equation}
    F(\bm{U}^{\text{dat}}, \dot{\bm{U}^{\text{dat}}}, \bm{U}^{\text{dat'}}, \cdots)=0
\end{equation}

Raissi et al.\cite{Raissi2019-ux} proposed physics-informed neural networks (PINNs) that output solutions that strictly satisfy the equations by introducing a constraint imposed by the governing equations as the loss function of the neural network. Samaniego et al.\cite{Samaniego2020-po} demonstrated that neural networks can solve various problems in solid mechanics by considering the functional of a variational problem equivalent to a differential equation as a loss function. Additionally, Liao et al.\cite{Liao2020-yu} organized a solution process for inverse analysis in materials engineering using a neural network constrained by physical conditions to design an atomic structure with desirable material properties.

Note that the method proposed in these studies differs from a simple surrogate model that naively learns and reproduces the input-output relationship of the simulated solution in that the solution of the data-driven model is explicitly constrained to satisfy the governing equations.

\subsection{Physics-based loss functions}
When using data-driven models for prediction and inverse analysis of physical phenomena, consistency with physical theories and lack of training data can be problematic. Therefore, methods have been proposed to constrain the solution space of data-driven models by imposing physical constraints in loss functions, to ensure physical consistency and to improve learning efficiency from a small amount of training data. The solutions obtained using these methods are expressed as follows:

\begin{equation}
    \bm{U}^{\text{hybrid}}=\bm{U}^{\text{dat}}|F(\bm{U}^{\text{dat}}, \dot{\bm{U}^{\text{dat}}}, \bm{U}^{\text{dat'}}, \cdots)=0
\end{equation}

The method described in the previous subsection differs from the method described in this subsection in the following points. The method described in the previous subsection is a data-driven solver that solves the problems modeled by the governing equations. Here, the representation ability of the entire model depends on the governing equations, and the main purpose is to achieve a high computational speed. In contrast, the methods discussed in this subsection impose physical constraints on the solution of data-driven models. Here, it is expected that the flexible representational performance of the data-driven model will be realized, and the purpose is to ensure the physical consistency of the output, to improve the learning efficiency, and as a result, improve the prediction performance.

Read et al.\cite{Read2019-mp} proposed a neural network that imposes an energy conservation law on the computation results (process-guided deep learning) to predict the water temperature distribution in a lake based on information such as time and air temperature. In this model, the neural network is trained to reproduce the input-output relationship of real data, while the output water temperature distribution is constrained by the loss function to satisfy the energy conservation law.  It is shown that the proposed method always has a higher prediction accuracy than ordinary neural networks and physical models, particularly when the training data are small.
Karpatne et al.\cite{Karpatne2019-bb} also mentioned that applying physical constraints to machine learning models can address the lack of training data and correct labels that are widely highlighted in geoscience inverse problems. Thus, the method for imposing physical constraints on the data-driven model is a method that can take advantage of the representational performance of the data-driven model while gaining various benefits as described above by using physical knowledge.

\subsection{Data-driven construction of governing equations}
The use of governing equations to describe phenomena is a universal method for clarifying and predicting phenomena based on laws derived from deductive reasoning. However, some phenomena difficult to deduce the governing equation for, such as when the phenomenon is complex, or its behavior is governed by microscopic laws\cite{Khain2015-ca}. For such events, data-driven methods were proposed to determine the governing equations based on the observed data of the phenomenon. The laws obtained in this method can be described as follows.

\begin{equation}
    F^{\text{D-driven}}(\bm{u}, \dot{\bm{u}}, \bm{u}',\cdots) = 0
\end{equation}

Brunton et al.\cite{Brunton2016-bv,Rudy2017-hk} proposed a method for expressing the governing equations as a linear sum of nonlinear functions of physical quantities called library functions and constructing the governing equations by identifying the main terms through a regression problem with sparse regularization. Additionally, Berg et al.\cite{Berg2019-rw} proposed a method that extends the linear sum in Brunton et al.'s method to a nonlinear transformation using a neural network. Furthermore, Koopman operator analysis is gaining traction as a data-driven method for determining the governing equations in a more general form without requiring the equations to be in the form of linear sums\cite{Brunton2016-jv}.
Koopman operator analysis is a method of reproducing the modal decomposition form, which is an equivalent representation of the time evolution law of dynamic phenomena from observed data, and has been studied mainly in the research field of dynamical systems\cite{Rowley2009-nc,Kevrekidis2016-dq,Morton2019-xe,Yeung2019-vk}. These methods are used for knowledge discovery, which is a data-driven attempt to discover the laws that the observed data must obey, and for building models that have extrapolation ability based on the laws.

Additionally, mainly for speeding up the computation, the reduced-order model, which obtains an approximate reduced-dimensional representation of the physical model from the data, has been proposed using proper orthogonal decomposition\cite{Volkwein2013-bf}, dynamic mode decomposition\cite{Proctor2016-kx}, and autoencoder\cite{Lee2020-ao}, and applied in fields such as numerical fluid dynamics analysis\cite{Hijazi2020-qn}.

\subsection{Reflecting physical knowledge in model architectures}
Most machine learning models, such as neural networks, have a black box nature, which makes it difficult to interpret the computational process and determine its validity. Additionally, the models used for deep learning have numerous parameters, and it is difficult to search for the optimal value from a huge solution space. Therefore, methods have been proposed to improve the interpretability of the computational process and the performance of learning and prediction by introducing physical knowledge into the architecture of data-driven models. Here, we describe a data-driven model with such physical knowledge as follows:

\begin{equation}
    \bm{U}^{\text{hybrid}} = f^{\text{phy-informed}}(\bm{x}, t;\bm{w})
\end{equation}

Mulalindhar et al.\cite{Muralidhar2020-ew} demonstrated that in designing a convolutional neural network to predict fluid properties, a network architecture in which the intermediate layers output the velocity and pressure fields of the fluid (physics guided neural networks) improve the interpretability of the computational process of the model, and demonstrated that it can achieve higher prediction performance than the physical model even when there is an insufficient amount of training data.
Furthermore, Greydanus et al.\cite{greydanus2019hamiltonian} demonstrated that when predicting the behavior of a dynamical system, the total energy of the system is conserved and the dynamic properties of the model output does not break down by using Hamiltonian neural networks that learn the Hamiltonian instead of learning the behavior itself, which improves the properties and performance of the model by devising the intermediate and output quantities of the data-driven model. Furthermore, Chen et al.\cite{Chen2018-ut} demonstrated that by replacing the discrete layer structure of neural networks with a continuous representation using differential equations, a neural network model (neural ODE) with superior computational efficiency and output value continuity can be constructed.

\subsection{Coupled analysis of physical and data-driven models}
Because physical and data-driven models have contradictory advantages and challenges, each method has its own suitability and unsuitability depending on the physical nature of the target analysis and of the data. Therefore, an approach has been devised in which physical models are used to analyze physical quantities suitable for physical analysis, while data-driven models are used to analyze physical quantities that are expected to benefit from data-driven modeling, and the two solutions are coupled.
Here, the solutions $\bm{U}^{\text{phy}}$ of the physical model and $\bm{U}^{\text{dat}}$ of the data-driven model are used to construct the overall prediction $\bm{U}^{\text{hybrid}}$ by referring to each other in the solution process.

\begin{equation}
    \bm{U}^{\text{hybrid}} = [\bm{U}^{\text{phy}}|\bm{U}^{\text{dat}}, \bm{U}^{\text{dat}}|\bm{U}^{\text{phy}}]
\end{equation}

Parish et al.\cite{Parish2016-vu} proposed a method to represent the difference between physical models and real phenomena in the governing equations using a Gaussian process, and applied it to turbulence models. Rasp et al.\cite{Rasp2018-tj} and Yuval et al.\cite{Yuval2020-zt} demonstrated that replacing the parameterization term of the subgrid process in a meteorological model with a machine learning model improves the accuracy of the analysis. Zheng et al.\cite{Zheng2020-nh} proposed and demonstrated an effective method for analyzing the atmospheric velocity field using kinematic model and rainfall variability using data-driven model to predict rainfall.
These methods can be described as complementary approaches to modeling phenomena that use both physical and data science methodologies to obtain more accurate models than those that use each methodology alone.

\section{Discussion}
The methods integrating the physical and data science methodologies mentioned so far have achieved results in overcoming the challenges of each methodology. In this section, we discuss the role of these results in achieving the two objectives of {\it prediction} and {\it inference} and discuss their future development.

James et al.\cite{James2013-sh} summarized the purpose of building a statistical machine learning model $\bm{y}=f(\bm{x})$ that describes the relationship between the input $\bm{x}$ and output $\bm{y}$ into two categories: {\it prediction} and {\it inference.} The accuracy of the estimated $\bm{y}$ is more important than the interpretability and validity of model $f$ in {\it prediction}, and  problem-solving and engineering perspectives are emphasized. {\it Inference}, on the other hand, focuses on understanding the relationship between x and y and the mechanism through model $f$. When {\it inference} is the objective, the validity and interpretability of model $f$ are essential, and knowledge discovery and a scientific perspective are emphasized. These two objectives are not limited to machine learning models but can be widely applied to mathematical models used for clarification, prediction, and control of phenomena in general.

\subsection{Prediction}
The integration methods mentioned in this study have made significant contributions to {\it prediction}. The methods proposed by Read et al.\cite{Read2019-mp}, Kubo et al.\cite{Kubo2020-fg}, Mulalindhar et al.\cite{Muralidhar2020-ew}, and Rasp et al.\cite{Rasp2018-tj} directly improved the prediction accuracy of the physical quantities of interest, whereas the methods proposed by Karpatne et al.\cite{Karpatne2017-uo} and Jia et al.\cite{Jia2021-os} allowed us to leverage on the flexible prediction capabilities of data-driven models even when sufficient labeled data were not available. Additionally, the computational speedup achieved by Zhang et al.\cite{Zhang2020-uk}, Ichimura et al.\cite{Ichimura2018-xp}, and Hijazi et al.\cite{Hijazi2020-qn} extended the scope of analysis and prediction. Furthermore, concrete applications of these methods are also underway. For example, Schmidt et al.\cite{Schmidt2019-ss} argued that by replacing part of the computational process with surrogate models, various practical processes in materials science can be accelerated, such as the discovery of desirable material structures and the prediction of physical properties.

These methods seem to work effectively in both directions: extending both the physical and data science methodologies. They improve model accuracy and speed up the computation for physical models, and they expand the range of applications for data-driven models by adding physical consistency and extrapolation capabilities.
From the perspective of contributing to the {\it prediction}, we expect that the proposal of new methods and the expansion of social implementation will continue to progress steadily in the future.

\subsection{Inference}
{\it Inference}, which is the discovery of mechanisms inherent in data, can be described as knowledge discovery from data. In the inverse problem of estimating the parameters of the governing equations from observed data, the methods presented by Gafeira et al.\cite{Gafeira2021-ad}, Raissi et al.\cite{Raissi2019-ux} and Karpatone et al.\cite{Karpatne2019-bb} provide fast and accurate methods for solving the problem. Rasp et al.\cite{Rasp2018-tj} and Yuval et al.\cite{Yuval2020-zt} used data-driven methods to construct parameterization terms in equations, which identify parts of the governing equations from data.
Furthermore, knowledge discovery from data continues to be one of the mainstream topics in machine learning and data mining, and Fayyad et al.\cite{Fayyad1996-sv} defined knowledge discovery as pattern discovery from data.
From this perspective, the reduced-order models are not only significant for speeding up the computation, but also for data-driven discovery of the dominant modes of physical quantities, a method that is significant for knowledge discovery.
These methods assume that the form of the governing equation of interest is available and identify some parameters in the equation or patterns in the solutions of the equation. As mentioned above, many integration methods have been successfully applied for this purpose.

In contrast, the discovery of the governing equations themselves, that is, the discovery of the laws inherent in the phenomena and data of interest, leads to a more principled elucidation of the mechanisms of phenomena. Although the discovery of universal laws from big data is widely expected in fields such as earth science\cite{Karpatne2019-bb}, materials science\cite{Schmidt2019-ss,Butler2018-bx} and fluid mechanics\cite{Witherden2017-or}, such efforts have had limited success.
Brunton et al.\cite{Brunton2016-bv} and Berg et al.\cite{Berg2019-rw} used a sum of pre-prepared nonlinear functions to find the governing equations, but it is assumed that the governing equations exist in the space spanned by the nonlinear functions.  The Koopman operator analysis method\cite{Rowley2009-nc,Kevrekidis2016-dq,Morton2019-xe,Yeung2019-vk} can be regarded as a method for determining the governing equations in a more general form, but it is known that it is difficult to apply it to phenomena involving spatial movement of physical quantities, such as wave fields and advection.

However, the method proposed by Lu et al.\cite{Lu2020-aa} and Zheng et al.\cite{Zheng2020-nh} enables the application of Koopman operator analysis to phenomena with advection by transforming data coordinates based on physical knowledge, and the integration of physics and data science remains a clue to the solution to the above problem of law discovery. For such an approach to be realized, a deep understanding of both methodologies is necessary, and the exchange of knowledge and the development of the field through interdisciplinary research and communication among experts is expected.

\section{Conclusion}
Methods that integrate physical and data science methodologies can address the challenges of each methodology by complementary means, increasing the options for knowledge discovery and problem solving. The history of paradigm shifts in science indicates that these methods are expected to function as a new scientific framework for clarifying, predicting, and controlling phenomena, and are being applied in various fields.

The linkage of data collection and informatics technologies, which is not the subject of this study, can also function together with modeling technologies as two wheels for developing science and society. From the perspective of prediction and control of phenomena, methodologies provided by research areas such as uncertainty quantification\cite{Wang2016-bg} and data assimilation\cite{Clark_Di_Leoni2020-qo} are important, and there is still room for further integration with the modeling methods outlined in this study. The shift and deepening of the scientific paradigm through data science is currently underway, and further development through systematic organization and understanding of the technologies is expected in the future.

\section*{Acknowledgment}
The author would like to express my gratitude to Dr. Marlon Nuske in German Research Center for Artificial Intelligence for his suggestive advice.

\bibliographystyle{unsrt}  
\bibliography{all}

\end{document}